\documentstyle[aps,epsf]{revtex}

\begin{document}

\title{Bianchi Type I Universes with Causal Bulk Viscous Cosmological Fluid}
\author{M. K. Mak\footnote{E-mail:mkmak@vtc.edu.hk}}
\address{Department of Physics, The Hong Kong University of Science and Technology,Clear Water Bay, Hong Kong, P. R. China.}
\author{T. Harko\footnote{E-mail: tcharko@hkusua.hku.hk}}
\address{Department of Physics, The University of Hong Kong,
Pokfulam Road, Hong Kong, P. R. China.}
\maketitle

\begin{abstract}

We consider the dynamics of a causal bulk viscous cosmological
fluid filled flat constantly decelerating Bianchi type I space-time. The
matter component of the Universe is assumed to satisfy a linear barotropic
equation of state and the state equation of the small temperature Boltzmann
gas. The resulting cosmological models satisfy the condition of smallness of
the viscous stress. The time evolution of the relaxation time, temperature,
bulk viscosity coefficient and comoving entropy of the dissipative
cosmological fluid is also obtained.


\end{abstract}


\section{Introduction}

In a series of recent papers \cite{BrPe97} -\cite{CaCa00}, a viscous
Bianchi type I Universe with metric of Kasner form has been investigated.
The cosmic fluid was supposed to be endowed with a bulk viscosity $\xi $ and
a shear viscosity $\eta $, with both viscosity coefficients independents of
position. The corresponding energy-momentum tensor is $T_{ik}=\rho
u_{i}u_{k}+\left( p-\xi \theta \right) h_{ik}-2\eta \sigma _{ik}$ , with $%
u_{i}\ $the four -velocity, $h_{ik}=g_{ik}+u_{i}u_{k}$ the projection tensor
, $\theta =u_{;i}^{i}$ the scalar expansion and $\sigma _{ik}=\frac{1}{2}%
\left( u_{i;l}h_{k}^{l}+u_{k;l}h_{i}^{l}\right) -\frac{1}{3}h_{ik}\theta $
the shear tensor. This energy-momentum tensor is of Eckart-Landau
(non-causal) type \cite{Ec40}, \cite{LaLi87}. 
The entropy current four-vector is given, in these type of models and in the absence of the heat flux, by $%
S^{i}=nk_{B}\sigma u^{i}$ ,where $k_{B}$ is the Boltzmann constant, $n$ is
the baryon number density and $\sigma $ the nondimensional entropy per
baryon. But  as has been recently shown in \cite{CaCa00}, if the thermodynamics together with the dominant energy
conditions are kept, the viscous cosmological fluid model in a Kasner type
geometry cannot describe the growth of the entropy in the Universe.

In many cosmological and astrophysical situations an idealised fluid model
of the matter is inappropriate. Such possible situations are the
relativistic transport of photons, mixtures of cosmic elementary particles,
the evolution of cosmic strings due to their interaction with each other and
with the surrounding matter, the classical description of the (quantum)
particle production phase, interaction between matter and radiation, quark
and gluon plasma viscosity, different components of dark matter etc. \cite{ChJa96}. From a physical point of view the inclusion of dissipative terms
in the energy momentum tensor of the cosmological fluid seems to be the
best-motivated generalisation of the matter term of the gravitational field
equations.

The theories of dissipation in Eckart-Landau formulation \cite{Ec40}, \cite{LaLi87}, who made the first attempt at creating a
relativistic theory of viscosity, are now known to be pathological in
several respects. Regardless of the choice of equation of state, all
equilibrium states in these theories are unstable. In addition, as shown by
Israel \cite{Is76}, signals may be propagated through the fluid at
velocities exceeding the speed of light. These problems arise due to the
first order nature of the theory since it considers only first-order
deviations from the equilibrium, leading to parabolic differential
equations, hence to infinite speeds of propagation for heat flow and
viscosity, in contradiction with the principle of causality. While such
paradoxes appear particularly glaring in relativistic theory, infinite
propagation speeds already constitutes a difficulty at the classical level,
since one does not expect thermal disturbances to be carried faster than
some (suitably defined) mean molecular speed. Conventional theory is thus
applicable only to phenomena which are ''quasi-stationary'' i.e. slowly
varying on space and time scales characterised by mean free path and mean
collision time \cite{Is76}. This is inadequate for many phenomena in
high-energy astrophysics and relativistic cosmology involving steep
gradients or rapid variations. These deficiencies can be traced to the fact
that the conventional theories (both classical and relativistic) make overly
restrictive hypothesis concerning the relation between the fluxes and
densities of entropy, energy and particle number.

A relativistic second-order theory was found by Israel \cite{Is76} and
developed by Israel and Stewart \cite{IsSt76} into what is called
'transient' or 'extended' irreversible thermodynamics. In this model
deviations from equilibrium (bulk stress, heat flow and shear stress) are
treated as independent dynamical variables leading to a total of $14$
dynamical fluid variables to be determined. The solutions of the full causal
theory are well behaved for all times. Hence the advantages of the causal
theories are the followings \cite{AnPaRo98}: 1) for stable fluid
configurations the dissipative signals propagate causally 2) unlike
Eckart-type's theories, there is no generic short-wavelength secular
instability in causal theories and 3) even for rotating fluids, the
perturbations have a well-posed initial value problem. Therefore, the best
currently available theory for analysing dissipative processes in the
Universe is the full Israel-Stewart causal thermodynamics.

Due to the complicated nonlinear character of the evolution equations, very
few exact cosmological solutions of the gravitational field equations are
known in the framework of the full causal theory. For a homogeneous Universe
filled with a full causal viscous fluid source obeying the relation $\xi
\sim \rho ^{\frac{1}{2}}$, exact general solutions of the field equations
have been obtained in \cite{ChJa97} and \cite{MaHa98} - \cite{MaHa99a}. In this
case the evolution of the bulk viscous cosmological model can be reduced to
a Painleve-Ince type differential equation. It has also been proposed that
causal bulk viscous thermodynamics can model on a phenomenological level
matter creation in the early Universe \cite{MaHa98}, \cite{MaHa99a}.

Because of technical reasons, most investigations of dissipative causal
cosmological models have assumed FRW symmetry (i.e. homogeneity and
isotropy) or small perturbations around it \cite{MaTr97}. The Einstein
field equations for homogeneous models with dissipative fluids can be
decoupled and therefore are reduced to an autonomous system of ordinary
differential equations, which can be analysed qualitatively \cite{CoHo95}, \cite{CoHo96}.
A qualitative analysis of the Bianchi type I
cosmological model has been performed, under the assumption of the plane
symmetry, by Romano and Pavon \cite{RoPa93}. Homogeneous and isotropic cosmological
fluids with bulk viscous dissipation have been analyzed, in the framework of the
truncated Israel-Stewart theory, with the use of dynamical systems techniques, by
Di Prisco, Herrera and Ibanez \cite{PrHeIb01}. They found that almost all solutions inflate and
only few of them can be considered physical, since the dominant energy condition is not satisfied.

The matter density parameter $\Omega $ and the deceleration parameter $q$
are the most important observational parameters in cosmology. The values of
the deceleration parameter separates decelerating ($q>0$ ) from accelerating
( $q<0$) periods in the evolution of the Universe. Determining the deceleration
parameter from the count magnitude relation for galaxies is a difficult task due to
evolutionary effects. The tightest limits on
the present value $q_{0}$ of the deceleration parameter obtained from
observations are $-1.27\leq q_{0}\leq 2$ \cite{KlGr86}. Studies of
galaxy counts from redshift surveys provide a value of $q_0=0.1$, with an
upper limit of $q_0<0.75$ \cite{KlGr86}.
Recent observations show that the deceleration parameter of the Universe $q$
is in the range $-1\leq q<0$, and the present-day Universe undergoes an
accelerated expansionary evolution \cite{Ri98}-\cite{Pe99}. But of course these
results do not exclude the existence of a decelerating phase in the early history
of our Universe.

Cosmological models with constant deceleration parameter have been intensively
investigated in the physical literature. It has been shown by Berman and
Gomide \cite{BeGo98} that all the phases of the Universe, i.e.,
radiation and pressure-free phase, may be considered as
particular cases of the $q=constant$ type. Incidentally, most of the
well-known perfect fluid models in general relativity and Brans-Dicke theory
are models with constant deceleration parameter.

It is the purpose of the present paper to consider causal bulk viscous
cosmological models in a flat Bianchi type I space-time. We assume that
during the cosmic evolution the deceleration parameter is constant and
positive. Thus we extend to the anisotropic case the previous investigation
of isotropic flat and  homogeneous constantly decelerating bulk viscous
cosmological models \cite{HaMa00}.
This model satisfies the condition of smallness of the viscous
stress. We also analyse the evolution of the physical parameters of the
dissipative cosmological fluid such as relaxation time, temperature, bulk
viscosity coefficient and comoving entropy. For the equilibrium pressure we
shall adopt a linear barotropic equation of state and the state equation of
the small temperature Boltzmann gas.

The present paper is organised as follows. The basic equations describing
the dynamics of a decelerating causal bulk viscous cosmological model are
obtained in Section II. The case of a viscous cosmological fluid obeying a
barotropic equation of state is considered in Section III. Boltzmann gas
filled viscous Bianchi type I Universes are discussed in Section IV. In
Section V we discuss and conclude our results.

\section{Thermodynamics, Field Equations and Consequences}

The energy momentum tensor of a relativistic fluid, with bulk viscosity as
the only dissipative phenomenon is
\begin{equation}\label{1}
T_{i}^{k}=\left( \rho +p+\Pi \right) u_{i}u^{k}-\left( p+\Pi \right) \delta
_{i}^{k}. 
\end{equation}

In Eq. (\ref{1}) $\rho $ is the energy density, $u^{i}$ ,$i=0,1,2,3$ , is the
four-velocity, ( $u_{i}u^{i}=1$), $\ p$ is the equilibrium pressure and $%
\Pi $ is the bulk viscous pressure. The particle flow vector is given by
$N^{i}=nu^{i}$, where $n\geq 0$ is the particle number density. $%
N_{;i}^{i}=0$ leads to the particle number conservation equation $\dot{n}%
+3nH=0$, with $H=\frac{1}{3}u_{;i}^{i}$ the Hubble factor (a semicolon
denotes the covariant derivative with respect to the metric). The
conservation law $T_{i;k}^{k}=0$ implies
\begin{equation}\label{2}
\dot{\rho}+3H\left( \rho +p\right) =-3H\Pi. 
\end{equation}

In the framework of causal thermodynamics and limiting ourselves to
second-order deviations from equilibrium, the entropy flow vector $S^{i}$
takes the form \cite{Ma95}:
\begin{equation}\label{3}
S^{i}=sN^{i}+\frac{R^{i}\left( N^{i},T_{i}^{k}\right) }{T}=sN^{i}-\frac{\tau
\Pi ^{2}}{2\xi T}u^{i}, 
\end{equation}
where $s$ is the entropy per particle, $\tau $ is the relaxation time, $T$
is the temperature and $\xi $ is the coefficient of bulk viscosity.

The causal evolution equation for the bulk viscous pressure $\Pi 
$ is given by \cite{Ma95}
\begin{equation}\label{4}
\tau \dot{\Pi}+\Pi =-3\xi H-\frac{1}{2}\tau \Pi \left( 3H+\frac{\dot{\tau}}{%
\tau }-\frac{\dot{\xi}}{\xi }-\frac{\dot{T}}{T}\right). 
\end{equation}

Equation (\ref{4}) arises as the simplest way (linear in $\Pi $) to satisfy the $H$
theorem (i.e., for the entropy production to be non-negative, $S_{;i}^{i}=%
\frac{\Pi ^{2}}{\xi T}\geq 0$) \cite{Is76}, \cite{Ma95}. The
Israel-Stewart theory \cite{IsSt76} is derived under the assumption that
the thermodynamic state of the fluid is close to equilibrium, which means
that the non-equilibrium bulk viscous pressure should be small when compared
to the local equilibrium pressure, that is $\left| \Pi \right| <p$. If this
condition is violated then one is effectively assuming that the linear
theory holds also in the non-linear regime far from equilibrium but for a
fluid description of the matter, the condition ought to be satisfied. Due to
the non-thermalizing character of dissipative expansion which require that
the rate of interactions maintaining equilibrium remains lower than the
expansion rate we have the following consistency condition on causal viscous
cosmologies: $\frac{1}{\tau }<H$.

We assume equations of state in the general form $\rho =\rho \left(
n,T\right), p=p\left( n,T\right)$, according to which the particle number
density $n$ and the temperature $T$ are the basic thermodynamical variables.
In this case one finds the following evolution law for the temperature of a
causal bulk viscous cosmological fluid
\begin{equation}\label{5}
\frac{\dot{T}}{T}=-3H\left( b+\frac{a\Pi }{\rho }\right), 
\end{equation}
where we denoted $b\equiv \left( \partial p/\partial T\right) /\left(
\partial \rho /\partial T\right) $ , $a=\rho /\left( T\partial \rho
/\partial T\right) $ and $\gamma =\left( \rho +p\right) /\rho $.

The growth of the total comoving entropy over a proper time interval $\left(
t_{0},t\right) $ is given by \cite{Ma95}
\begin{equation}\label{6}
S \left( t\right) -S \left( t_{0}\right) =-\frac{3}{k_{B}}%
\int_{t_{0}}^{t}\frac{\Pi VH}{T}dt, 
\end{equation}
where $V$ is the proper comoving volume element.

In the case of a flat Bianchi type I space-time with a line element
\begin{equation}\label{7}
ds^{2}=dt^{2}-a_{1}^{2}(t)dx^{2}-a_{2}^{2}(t)dy^{2}-a_{3}^{2}(t)dz^{2}, 
\end{equation}
the gravitational field equations take the form
\begin{equation}\label{8}
3\dot{H}+H_{1}^{2}+H_{2}^{2}+H_{3}^{2}=-\frac{1}{2}\left( \rho +3p+3\Pi
\right), 
\end{equation}
\begin{equation}\label{9}
\frac{1}{V}\frac{d}{dt}\left( VH_{i}\right) =\frac{1}{2}\left( \rho -p-\Pi
\right) ,i=1,2,3, 
\end{equation}
where we denoted $V=a_{1}a_{2}a_{3}$ , $H_{i}=\frac{\dot{a}_{i}}{a_{i}}%
,i=1,2,3$ and $H=\frac{1}{3}\sum_{i=1}^{3}H_{i}=\frac{1}{3}\frac{\dot{V}}{V}$
.

By adding Eqs. (\ref{9}) we obtain
\begin{equation}\label{10}
\dot{H}+3H^{2}=\frac{1}{2}\left( \rho -p-\Pi \right), 
\end{equation}
leading to
\begin{equation}\label{11}
H_{i}=H+\frac{K_{i}}{V},i=1,2,3, 
\end{equation}
where $K_{i},i=1,2,3$ are constants of integration satisfying the
consistency condition $\sum_{i=1}^{3}K_{i}=0$.

The energy density $\rho $ of of the bulk viscous cosmological fluid is
given by
\begin{equation}
\rho =3H^{2}\left( 1-\frac{K^{2}}{6H^{2}V^{2}}\right),
\end{equation}
where $K^{2}=\sum_{i=1}^{3}K_{i}^{2}$. The weak energy condition $\rho \geq
0$ imposes the constraint $K^{2}\leq 6H^{2}V^{2}$ on the integration
constants $K_{i}, i=1,2,3$.

As a first example of a Bianchi type I space-time we consider the Kasner
type Universe for which the components of the metric tensor are given by $%
a_{i}(t)=t^{p_{i}}, i=1,2,3$ and $p_{i}=const.$. We denote $%
M=\sum_{i=1}^{3}p_{i}$. Hence we have $V=t^{M}$ , $H_{i}=\frac{p_{i}}{t}%
,i=1,2,3$ and $H=\frac{M}{3t}$. The coefficients $p_{i},i=1,2,3$ must
satisfy the consistency conditions, which follows from $\frac{1}{V}\frac{d}{%
dt}\left( VH_{i}\right) =\dot{H}+3H^{2},i=1,2,3$ and are given by $%
3p_{i}=p_{1}+p_{2}+p_{3},i=1,2,3$. These equations represent a homogeneous
system of algebraic equations. Since the determinant of the system is zero,
it admits a unique non-trivial solution given by $p_{1}=p_{3},p_{2}=p_{3}$.
Consequently, it follows that there are no anisotropic Kasner-type causal
bulk viscous cosmological fluid filled space-times. On the other hand this result is generally
valid for all cosmological fluids with isotropic pressure distribution.

A useful observational quantity in cosmology is the deceleration parameter
\begin{equation}\label{12}
q=\frac{d}{dt}\left( \frac{1}{H}\right) -1=\frac{\rho +3p+3\Pi +2\frac{K^{2}%
}{V^{2}}}{2\left( \rho +\frac{K^{2}}{2V^{2}}\right) }, 
\end{equation}
The positive sign of $q$ corresponds
to standard decelerating models whereas the negative sign indicates
inflation.

For a barotropic ($p=\left( \gamma -1\right) \rho ,\gamma =const.$ and $%
1\leq \gamma \leq 2$) cosmological perfect fluid filled flat isotropic FRW
type space -time the deceleration parameter is a constant for all $\gamma $
. In this case the gravitational field equations are $3H^{2}=\rho $, $2\dot{%
H}+3H^{2}=-\left( \gamma -1\right) \rho $, leading to $q=\frac{3\gamma }{2}%
-1$, respectively. The deceleration parameter is constant in both radiation
($\gamma =\frac{4}{3}$ ) and matter ($\gamma =1$ ) dominated phases. For a
barotropic perfect fluid filled Bianchi type I geometry using Eq.
(\ref{10}) the deceleration parameter can be expressed as 
\begin{equation}
q=2-\frac{\left( 2-\gamma \right) \rho }{2H^{2}}=\frac{3\gamma }{2}-1+\frac{%
9\left( 2-\gamma \right) }{4}\frac{K^{2}}{\dot{V}^{2}}.
\end{equation}

For $\gamma =2$ (stiff or Zeldovich fluid) $q$ is constant for all times.
For a rapidly expanding Universe with $\dot{V}>>K$, it is a very good
approximation to also consider the deceleration parameter as a constant ($q$ is
an exact constant for $\dot{V}=const.$ ). 

Since the bulk viscous pressure represents only a small correction to the
thermodynamical pressure, it is a reasonable assumption that the inclusion
of viscous terms in the energy -momentum tensor does not change
fundamentally the dynamics of the cosmic evolution. 
Therefore in the followings we restrict our analysis to cosmological evolutions
characterised by the constancy of the deceleration parameter,
\begin{equation}\label{13}
q=q_{0}=constant>0. 
\end{equation}

With this assumption the Hubble factor $H$ immediately follows from Eq. (\ref{12})
and is given by $H=\frac{H_{0}}{t}$, where $H_{0}=\frac{1}{q_{0}+1}=constant$. Hence the
volume factor is $V=V_{0}t^{3H_{0}}$, where $V_{0}>0$ is a constant of
integration. With the Hubble factor fixed by Eq. (\ref{13}) the time evolution of the
scale factors is given by
\begin{equation}\label{14}
a_{i}(t)=a_{i0}t^{H_{0}}e^{\frac{K_{i}}{V_{0}\left( 1-3H_{0}\right) }%
t^{1-3H_{0}}},i=1,2,3, 
\end{equation}
where $a_{i0},i=1,2,3$ are constants of integration.

From the field equations we obtain the energy density of the cosmological
fluid in the form
\begin{equation}\label{15}
\rho (t)=\frac{\rho _{0}}{t^{2}}\left( 1-\frac{C}{t^{6H_{0}-2}}\right), 
\end{equation}
where $\rho _{0}=3H_{0}^{2}$ and $C=\frac{K^{2}}{%
6H_{0}^{2}V_{0}^{2}}>0$. The sum of the thermodynamic and viscous pressure is
\begin{equation}\label{16}
p+\Pi =\frac{H_{0}\left( 2-3H_{0}\right) }{t^{2}}\left( 1-\frac{K^{2}}{%
2H_{0}\left( 2-3H_{0}\right) V_{0}^{2}}\frac{1}{t^{6H_{0}-2}}\right). 
\end{equation}

For $6H_{0}-2>0$, or, equivalently, $q_{0}<2$, the energy density $\rho $
of the cosmological fluid is positive only for time intervals $%
t>t_{0}=\left( \frac{K^{2}}{6H_{0}^{2}V_{0}^{2}}\right) ^{\frac{1}{6H_{0}-2}%
} $.

We also introduce the anisotropy parameter defined by
\begin{equation}
A=\frac{1}{3}%
\sum_{i=1}^{3}\left( \frac{H_{i}-H}{H}\right) ^{2}=\frac{K^{2}}{3V^{2}H^{2}}=\frac{K^{2}}{%
3V_{0}^{2}H_{0}^{2}}t^{2\left( 1-3H_{0}\right) }.
\end{equation}

The variation of the
particle number density of the bulk viscous cosmological fluid follows from
$n=\frac{n_{0}^{\prime }}{V}=\frac{n_{0}}{t^{3H_{0}}}$, where $%
n_{0}^{\prime }>0$ and $n_{0}>0$ are constants of integration.

The general solution of the gravitational field equations for a Bianchi type I space-time with causal bulk viscous fluid
can be generally expressed in an exact form for an arbitrary time dependent deceleration parameter $q=q(t)$. In the Appendix we present
the general representation of the solution in terms of the deceleration parameter.

\section{Models with a Linear Barotropic Equation of State}

In order to completely solve the gravitational field and the causal
evolution Eqs. (\ref{4}), (\ref{8}) and (\ref{9}), together with the condition of the
constancy of the deceleration parameter for an anisotropic Bianchi type I
geometry, we must close the system by an equation of state for $p$ and
specify the thermodynamic coefficients $\xi $ and $\tau $. We assume first
that the thermodynamic pressure of the bulk viscous cosmological fluid obeys
a linear barotropic equation of state 
\begin{equation}\label{17}
p(t)=\left( \gamma -1\right) \rho (t)=\frac{\left( \gamma -1\right) \rho _{0}%
}{t^{2}}\left( 1-\frac{C}{t^{6H_{0}-2}}\right),
\end{equation}
where $\gamma $ (the adiabatic index) is a constant and $1\leq \gamma \leq 2$%
. Hence the field equations (\ref{14}) and (\ref{15}) can be immediately solved to give:
\begin{equation}\label{18}
\Pi (t)=\frac{\Pi _{0}}{t^{2}}\left( 1-\frac{\left( 2-\gamma \right) C_{1}}{%
t^{6H_{0}-2}}\right), 
\end{equation}
where $\Pi _{0}=H_{0}\left( 2-3\gamma H_{0}\right) $ and $C_{1}=\frac{K^{2}}{%
2V_{0}^{2}\Pi _{0}}$.

Since for a viscous type dissipative theory it is expected that the bulk
viscous pressure is negative, $\Pi <0$, it follows that $2-3\gamma H_{0}<0$ leading to $q_{0}<\frac{%
3\gamma -2}{2}$. For $\gamma =\frac{4}{3}$ (radiation gas) we obtain $q_{0}<1$.

The causal evolution Eq. (\ref{4}) can be formally integrated to give
\begin{equation}\label{19}
T(t)=T_{01}\frac{\tau \Pi ^{2}}{\xi }\exp \left[ 2\int \left( \frac{1}{\tau }%
+\frac{3\xi H}{\tau \Pi }+\frac{3H}{2}\right) dt\right], 
\end{equation}
with $T_{01}$ a constant of integration. 

Following Belinskii, Nikomarov and Khalatnikov \cite{BeNiKh79}, we 
suppose that the relation $\tau =\frac{\xi }{\rho }$
holds in order to guarantee that the propagation velocity of bulk viscous
perturbations, i.e., the non-adiabatic contribution to the speed of sound in
a dissipative fluid without heat flux or shear viscosity, which is expected
to be of the order, $c_{b}=\left[ \xi /\left( \rho +p\right) \tau \right]
^{1/2}\sim \left[ \xi /\rho \tau \right] ^{\frac{1}{2}}$ does not exceed the
speed of light. With this choice for the relaxation time $\tau $ the causal
evolution Eq. (\ref{4}) gives the following evolution law for the temperature:
\begin{equation}\label{20}
T(t)=T_{0}t^{l}\left( 1-\frac{C}{t^{6H_{0}-2}}\right) ^{-1}\left[ 1-\frac{%
\left( 2-\gamma \right) C_{1}}{t^{6H_{0}-2}}\right] ^{m}e^{2\int \frac{\rho 
}{\xi }dt}, 
\end{equation}
where we denoted $T_{0}=T_{01}\frac{\Pi _{0}^{2}}{\rho _{0}}$ , $l=\frac{%
6\rho _{0}H_{0}}{\Pi _{0}}+3H_{0}-2$ and $m=2+\frac{6H_{0}^{2}}{\left(
2-\gamma \right) \Pi _{0}}$.

An analysis of the relativistic kinetic equation for some simple cases given
by Murphy \cite{Mu73}, Belinskii and Khalatnikov \cite{BeKh75} and
Belinskii, Nikomarov and Khalatnikov \cite{BeNiKh79} has shown that in
the asymptotic regions of small and large values of the energy density, the
viscosity coefficients can be approximated by power functions of the energy
density with definite requirements on the exponents of these functions. For
small values of the energy density it is reasonable to consider large
exponents, equal in the extreme case to one. For large $\rho $ the power of
the bulk viscosity coefficient should be considered smaller (or equal) to $%
1/2$. Hence, we shall assume that the bulk viscosity coefficient obeys the
simple phenomenological law
\begin{equation}\label{21}
\xi (t)=\alpha \rho ^{s}=\frac{\alpha \rho _{0}^{s}}{t^{2s}}\left( 1-\frac{C%
}{t^{6H_{0}-2}}\right) ^{s}, 
\end{equation}
with $\alpha $ and $s$ constants. Consequently, using Eq. (\ref{21}) the evolution
of the temperature in a bulk viscous cosmological fluid filled constantly
decelerating Bianchi type I Universe is given by
\begin{equation}\label{22}
T(t)=T_{0}t^{l}\left( 1-\frac{C}{t^{6H_{0}-2}}\right) ^{-1}\left[ 1-\left(
2-\gamma \right) \frac{C_{1}}{t^{6H_{0}-2}}\right] ^{m}F(t), 
\end{equation}
where $F(t)=\exp \left[ \frac{2}{\alpha \rho _{0}^{s-1}}\int \frac{%
t^{2(s-1)}dt}{\left( 1-\frac{C}{t^{6H_{0}-2}}\right) ^{s-1}}\right] $.

The evolution of the temperature for time intervals $t>C^{\frac{1}{6H_{0}-2}}
$ is presented, for a radiation dominated cosmological fluid ($\gamma =%
\frac{4}{3}$) in Fig.1.

\begin{figure}[h]
\epsfxsize=10cm
\centerline{\epsffile{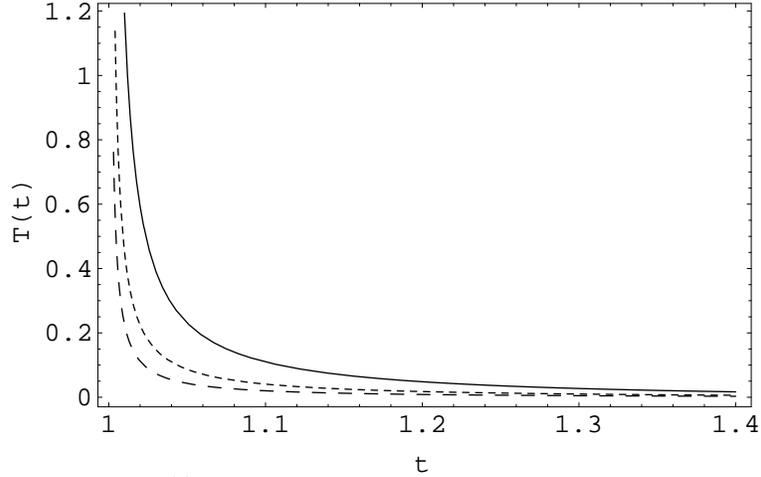}}
\caption{Behaviour of the temperature $T(t)$ of the barotropic causal bulk viscous cosmological fluid
for different values of the deceleration parameter:
$q_0=\frac{1}{4}$ (solid curve), $q_0=\frac{1}{2}$ (dotted curve) and $q_0=\frac{3}{4}$ (dashed curve).}
\label{FIG1}
\end{figure}

For all times and for  values of $q_{0}<2$ the temperature is a monotonically decreasing function of time. 

The behaviour of the comoving entropy is governed by Eq.(\ref{6}). Since the bulk
viscous pressure $\Pi <0$ , from Eq.(\ref{6}) it follows that $\frac{dS}{dt}%
\geq 0$. The variation of the entropy as a function of time for different
values of the deceleration parameter is presented in Fig.2.

\begin{figure}[h]
\epsfxsize=10cm
\centerline{\epsffile{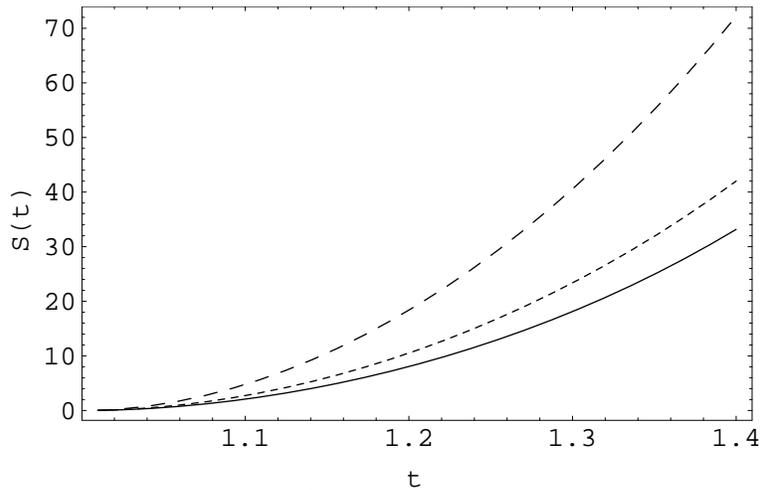}}
\caption{Time dependence of the comoving entropy $S(t)$ of the barotropic causal bulk viscous cosmological fluid for different values of the deceleration parameter:
$q_0=\frac{1}{4}$ (solid curve), $q_0=\frac{1}{2}$ (dotted curve) and $q_0=\frac{3}{4}$ (dashed curve).} 
\label{FIG2}
\end{figure}

The entropy is a monotonically increasing function of time for deceleration parameters in the
assumed range. Hence due to bulk viscous type dissipative stresses a large
amount of entropy can be produced during the very early evolution of our
universe. 

Finally, we shall check the condition of the smallness of the
non-equilibrium bulk viscous pressure $\Pi (t)$ with respect to the local
equilibrium pressure, $p(t)$, meaning that the thermodynamical state of the
fluid is close to equilibrium. From Eqs. (\ref{17}) and (\ref{18}) we obtain
\begin{equation}\label{23}
\left| \frac{\Pi (t)}{p(t)}\right| =\left| \frac{\Pi _{0}}{\left( \gamma
-1\right) \rho _{0}}\left( 1-\frac{C}{t^{6H_{0}-2}}\right) ^{-1}\left( 1-%
\frac{3(2-\gamma )CH_{0}}{2-3\gamma H_{0}}\frac{1}{t^{6H_{0}-2}}\right) \right|<1.
\end{equation}

The condition (\ref{23}) is satisfied for $q_{0}<2$ (in the small time limit) and $%
q_{0}<\frac{6\gamma -5}{2}$ (large time limit). For $\gamma =\frac{4}{3}$ we
obtain $q_{0}<\frac{3}{2}$. Therefore, the cosmological solutions obtained
above are thermodynamically consistent.

\section{Matter Dominated Causal Bulk Viscous Bianchi type I Cosmological Models}

In the previous Section we have considered that the bulk viscous
cosmological fluid filling the early universe satisfies a linear barotropic
equation of state, thus having a radiation-like behaviour. We consider
now the case of a bulk viscous  matter dominated cosmological fluid
consisting from a relativistic classical monatomic Boltzmann gas (i.e. a
dilute monatomic gas with high collision rate). By denoting $\beta =\frac{m}{%
k_{B}T}$ the ideal-gas law for the Boltzmann gas take the form: $p=nm\beta
^{-1}$, $nm=A_{0}e^{-\alpha }\beta ^{-1}K_{2}\left( \beta \right) $ and $%
\rho =A_{0}\left[ \beta ^{-1}K_{1}\left( \beta \right) +3\beta
^{-2}K_{2}\left( \beta \right) \right] $, with $\alpha $
the relativistic chemical potential,the constant $A_{0}=m^{4}g/2\pi \hbar
^{3}$, $g$ is the spin weight of the fluid particles and $K_{i},i=1,2,3$
are modified Bessel (Hankel) functions. For a Boltzmann
gas the relaxation time and the bulk viscosity coefficient can be expressed
as $\tau =1/\gamma _{0}n$ and $\xi =mc\eta ^{2}\Omega ^{2}/\gamma _{0}$,
with $\eta =K_{3}\left( \beta \right) /K_{2}\left( \beta \right) $, $\Omega
=3\gamma \left[ 1+1/\beta \eta \right] -5$ and $\gamma $ is a solution to $%
\left[ \gamma /(\gamma -1)\right] =\beta ^{2}\left( 1+\left( 5\eta /\beta
\right) -\eta ^{2}\right) $ \cite{Be88}. $\gamma _{0}$ is the collision
integral.

The equations governing a flat Robertson-Walker cosmological model
containing a dissipative Boltzmann gas have been integrated numerically
within the framework of the full causal dissipative thermodynamics by
Hiscock and Salmonson \cite{HiSa91}. In this case the evolution of the
Universe is non-inflationary and the pressure is not allowed to become
negative. Inhomogeneous Lemaitre-Tolman-Bondi type hydrodynamic models for a
cosmological matter fluid with causal viscous stresses but no heat
conduction satisfying the above equations of state have recently been
considered in \cite{Su97} - \cite{Su98}.

The equation of state of the Boltzmann gas in the form given above are not
especially enlightening. Hence it is more useful to evaluate the state
equation of the relativistic Boltzmann gas in two regimes: in the high
temperature limit corresponding to $\frac{T}{m}>>1$ and in the small
temperature limit for which the physical parameters of the gas satisfies the
condition $\frac{T}{m}<<1$.

In the small temperature limit the equation of state of the Boltzmann gas is
\cite{Be88}
\begin{equation}\label{24}
\rho =mn+\frac{3}{2}nT,P=nT,\frac{T}{m}<<1.
\end{equation}

For a matter fluid satisfying equations of state of the form (\ref{24}) we have $%
\gamma =1$, $a=(2m/3T)>>1$,and $b=(2/3)$. We assume that the gas
evolves in a flat anisotropic Bianchi type I space-time and the cosmological
evolution is characterised by a constant positive deceleration parameter, $%
q=q_0=constant>0$. With this assumption the energy density of the low temperature
Boltzmann gas is given again by Eq. (\ref{15}). Then from the gravitational field equations (\ref{5})-(\ref{6}) and with the use of 
Eqs. (\ref{15}) and (\ref{24}) we obtain:
\begin{equation}\label{25}
p(t)=\frac{2\rho _{0}}{3}\frac{1}{t^{2}}\left( 1-\frac{C}{t^{6H_{0}-2}}-%
\frac{mn_{0}}{\rho _{0}}\frac{1}{t^{3H_{0}-2}}\right), 
\end{equation}
\begin{equation}\label{26}
T(t)=\frac{2\rho _{0}}{3n_{0}}\frac{1}{t^{2-3H_{0}}}\left( 1-\frac{C}{%
t^{6H_{0}-2}}-\frac{mn_{0}}{\rho _{0}}\frac{1}{t^{3H_{0}-2}}\right), 
\end{equation}
\begin{equation}\label{27}
\Pi (t)=\frac{H_{0}\left( 2-5H_{0}\right) }{t^{2}}\left[ 1-\frac{CH_{0}}{%
2-5H_{0}}\frac{1}{t^{6H_{0}-2}}+\frac{2}{3}\frac{mn_{0}}{H_{0}\left(
2-5H_{0}\right) }\frac{1}{t^{3H_{0}-2}}\right].
\end{equation}

Equations (\ref{25})-(\ref{27}) are physically meaningful only for time intervals
satisfying the condition $1-\frac{C}{t^{6H_{0}-2}}-\frac{mn_{0}}{\rho _{0}}%
\frac{1}{t^{3H_{0}-2}}>0$, thus leading to positively defined thermodynamic
pressure and temperature.

The requirement that the temperature be a decreasing function of time leads
to the condition $2-3H_{0}>0$ or to the restriction $q_{0}>\frac{1}{2}$
imposed on the deceleration parameter. The condition of the negativity of
the bulk viscous pressure implies $q_{0}<\frac{3}{2}$.

The causal evolution equation (\ref{4}) for the bulk viscous pressure  can be
written in the form
\begin{equation}\label{28}
\frac{d}{dt}\left( \frac{\xi }{\tau }\right) =6\frac{H}{\Pi }\left( \frac{%
\xi }{\tau }\right) ^{2}+\left[ \frac{2}{\tau }+\frac{d}{dt}\ln \left( \frac{%
t^{3H_{0}}\Pi ^{2}}{T}\right) \right] \left( \frac{\xi }{\tau }\right), 
\end{equation}
leading to the following expression for $f=\frac{\xi }{\tau }$:
\begin{equation}\label{29}
f=\frac{\xi }{\tau }=%
\frac{t^{3H_{0}}\Pi ^{2}}{T}\frac{\exp \left( 2\int \frac{dt}{\tau }\right) 
}{C_{2}-6H_{0}\int \left[ \frac{t^{3H_{0}-1}\Pi }{T}\exp \left( 2\int \frac{%
dt}{\tau }\right) \right] dt},
\end{equation}
where $C_{2}$ is a constant of integration.

If the condition
$\left| 6\frac{\xi ^{2}}{\tau ^{2}}\frac{H}{\Pi }\right| >>\left| \frac{2\xi 
}{\tau ^{2}}\right| $
or, equivalently, $\left| \xi \right| >>\left| \frac{\Pi }{3H}\right| $is fulfilled, we can
neglect in Eq. (\ref{28}) the term $\frac{2\xi }{\tau ^{2}}$. In this case we
obtain for $\frac{\xi }{\tau }$ the following approximate form:
\begin{equation}\label{30}
\frac{\xi }{\tau }\approx \frac{t^{3H_{0}}\Pi ^{2}}{T}\left(
C_{2}-6H_{0}\int \frac{t^{3H_{0}-1}\Pi }{T}dt\right) ^{-1}.
\end{equation}

The variation of the bulk viscosity coefficient-relaxation time ratio is
represented in Fig.3.

\begin{figure}[h]
\epsfxsize=10cm
\centerline{\epsffile{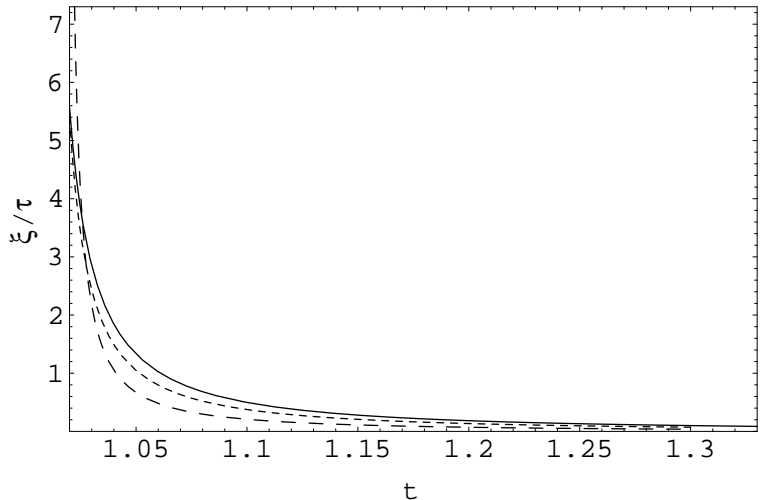}}
\caption{Time evolution of the bulk viscosity coefficient-relaxation time ratio $f(t)=\frac{\xi }{\tau }$
for the low temperature Boltzmann gas filled Bianchi type I space-time 
for different values of the deceleration parameter:
$q_0=0.6$ (solid curve), $q_0=0.8$ (dotted curve) and $q_0=1.1$ (dashed curve).} 
\label{FIG3}
\end{figure}

Another thermodynamic model can be obtained if we consider that the
relaxation time is related to the energy density of the low temperature
Boltzmann gas by means of the  relation $\frac{\xi }{\tau }=\rho $.

Then the causal evolution Eq. (\ref{4}) leads to the general relation
\begin{equation}\label{31}
\tau =-\frac{2}{\frac{18H^{3}}{\Pi }+\frac{d}{dt}\left[ \ln \left( \frac{%
t^{3H_{0}}\Pi ^{2}}{\rho T}\right) \right] }. 
\end{equation}

The behaviour of the relaxation time $\tau $ is presented, for different values of the
deceleration parameter, in Fig.4.

\begin{figure}[h]
\epsfxsize=10cm
\centerline{\epsffile{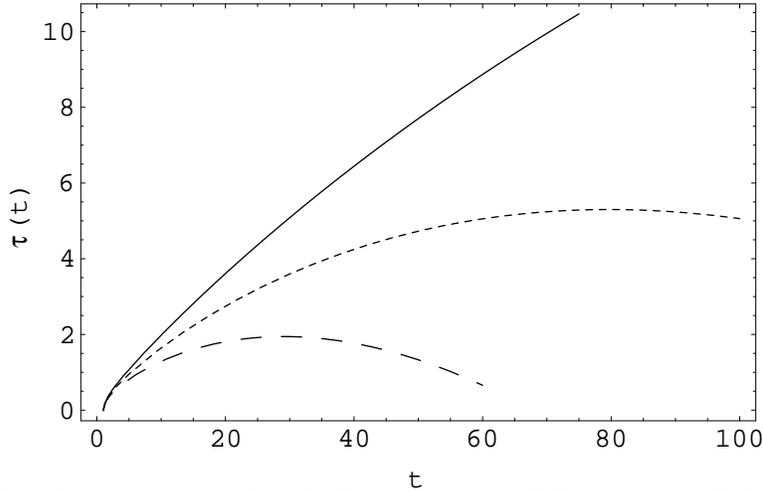}}
\caption{ Dynamics of the relaxation time $\tau (t)$ in the low temperature Boltzmann gas filled
Bianchi type I Universe for different values of the deceleration parameter:
$q_0=1.05$ (solid curve), $q_0=1.15$ (dotted curve) and $q_0=1.25$ (dashed curve).} 
\label{FIG4}
\end{figure}

In the large time limit, as expected, the relaxation time tends to zero.

The variation of the comoving entropy of a monatomic gas in a constantly
decelerating Universe can be obtained  from Eq. (\ref{6}). In Fig.5 we present
the time evolution of the entropy for the low temperature Boltzmann gas
filling an anisotropic constantly decelerating Bianchi type I Universe.  

\begin{figure}[h]
\epsfxsize=10cm
\centerline{\epsffile{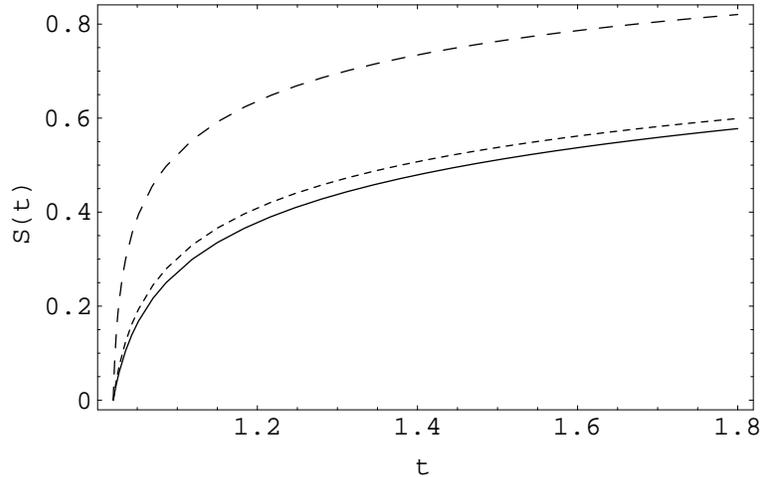}}
\caption{Variation of the entropy $S(t)$ of the low temperature Boltzmann gas filled
Bianchi type I Universe for different values of the deceleration parameter:
$q_0=0.6$ (solid curve), $q_0=0.8$ (dotted curve) and $q_0=1.1$ (dashed curve).} 
\label{FIG5}
\end{figure}

The condition of the smallness of the bulk viscous pressure can be studied
from the relation $\left| \frac{\Pi (t)}{p(t)}\right| <1$. In the limit of
large and small times we obtain the restrictions $q_{0}<\frac{5}{2}$ and $q_{0}<2$, respectively.
Hence the thermodynamic consistency of the low temperature dissipative
Boltzmann gas filled Bianchi type I  Universe is realised for values of the
deceleration parameter in the range $q_{0}\in \left( \frac{1}{2}, \frac{3}{2}%
\right) $.

The Boltzmann gas is not a physically realistic model of the content of the
Universe at all temperatures and densities (among other problems it ignores
the creation of particle-antiparticle pairs at relativistic temperatures).
But it is (together with the radiation fluid) the best physically motivated
and analysed cosmological fluid model. Even the low temperature approach is
still reasonable at high temperatures (for example, for the electron $%
m\approx 10^{9}K$ and the low temperature limit equations of state should be
very accurate up to about $10^{6}K$).

\section{Discussions and Final Remarks}

The study of the evolution and dynamics of the early Universe is usually
performed by looking for the consequences of the applications of the
physical laws to the cosmological environment. In the present paper we have
used the inverse approach, by investigating the effects of the cosmology on
the laws of the physics. At least for small (on a cosmological scale)
intervals of time  the deceleration parameter can be considered a constant
or a slowly varying parameter. The conditions of the constancy and of the
non-negativity of $q$ impose strong constraints on the evolution of the
cosmological models. For causal bulk viscous cosmological models these
restrictions lead, once we adopt an equation of state for the dense matter
and via the gravitational field equations, to explicit expressions for the
energy density, thermodynamic pressure and bulk viscous pressure. In the
present paper we have considered two distinct models, depending on the
equation of state of the cosmological fluid and on the relation. In all
these cases explicit expressions for the time dependence of the
thermodynamic parameters (temperature, bulk viscosity coefficient and
relaxation time) can be obtained. In these physical models the thermodynamic
consistency of the results requires values of $q$ so that $q<\frac{3}{2}$,
which is consistent with the range of values of the deceleration parameter
obtained from observations \cite{KlGr86}. 

In the large time limit, $t\rightarrow \infty $ \ and if $%
1-3H_{0}<0$ or, equivalently, $q_{0}<2$ , from Eqs. (\ref{14}) it follows that  $%
a_{i}(t)\rightarrow t^{H_{0}},i=1,2,3$ and the mean anisotropy parameter $%
A\rightarrow 0$. Hence due to the presence of dissipative stresses a
Bianchi type I universe ends in an isotropic phase. The transition from the anisotropic to the isotropic
state is associated to a large increase in the entropy. The increase of the comoving entropy
of the cosmological fluid is independent, in the causal thermodynamical approach,
of the equation of state of the matter and seems to be a general feature of the model. Hence this approach
can lead to a better understanding of the entropy generation mechanisms
in the very early Universe, where viscous type physical processes have probably played an important role
even when the Universe was about $1000s$ old.

Therefore the constantly decelerating causal bulk viscous cosmological
fluid model in an anisotropic geometry  leads to a self-consistent
thermodynamic description which maybe could describe a well-determined
period of the evolution of our Universe.

\section*{Appendix}

In this Appendix we present the general solution of the gravitational field 
equations for a causal bulk viscous cosmological fluid filled Bianchi type
I Universe in the case of a known functional form of the deceleration
parameter $q=q(t)$. From the definition of $q$ we obtain first
\begin{equation}
H=\left( q_{1}+t+\int q(t)dt\right) ^{-1},
\end{equation}
with $q_{1}$ a constant of integration (in the main text we have obtained all the
results with the particular choice $q_{1}=0$).

With the use of $3HV=\dot{V}$ the volume scale factor can be represented as 
\begin{equation}
V=V_{0}\exp \left[ 3\int \frac{dt}{q_{1}+t+\int q(t)dt}\right],
\end{equation}
with $V_{0}$ a constant of integration. The scale factors are
\begin{equation}
a_{i}=a_{i0}\exp \left[ \int \frac{dt}{q_{1}+t+\int q(t)dt}+\frac{K_{i}}{%
V_{0}}\int \frac{dt}{\exp \left( 3\int \frac{dt}{q_{1}+t+\int q(t)dt}\right) 
}\right] ,i=1,2,3,
\end{equation}
with $a_{i0},i=1,2,3$ arbitrary constants.

From Eqs.(\ref{8}) and (\ref{10}) we obtain the energy density of the bulk viscous fluid
in the general form
\begin{equation}
\rho =3H^{2}-\frac{K^{2}}{2V^{2}}=3H^{2}\left( 1-\frac{A(t)}{2}\right). 
\end{equation}

We assume that the cosmological fluid obeys the linear barotropic equation
of state $p=\left( \gamma -1\right) \rho $. Then the general form of the bulk viscous pressure
follows from the gravitational field equations and is given by
\begin{equation}
\Pi =\rho -p-2\left( \dot{H}+3H^{2}\right) =\frac{\left( \gamma -2\right)
K^{2}}{2V^{2}}-3\gamma H^{2}-2\dot{H}<0
\end{equation}

The condition of the negativity of the bulk viscous pressure is satisfied
only if the deceleration $q(t)$ and the mean anisotropy $A(t)$ parameters
obey the inequality
\begin{equation}
2\left[ 1+q(t)\right] -3\gamma <\frac{3}{2}\left( 2-\gamma \right) A(t).
\end{equation}

From Eq. (\ref{19}) and with the use of the evolution laws $\tau =\alpha \rho
^{s-1}$ and $\xi =\alpha \rho ^{s}$ for the relaxation time and bulk
viscosity coefficient respectively, we obtain the evolution law of the
temperature in the form
\begin{equation}
T(t)=T_{01}\frac{\left[ (2-\gamma )\rho -2\left( \dot{H}+3H^{2}\right)
\right] ^{2}}{\rho }
\exp \left\{ 2\int \left[ \frac{\rho ^{1-s}}{\alpha }%
+3H\left( \frac{\rho }{(2-\gamma )\rho -2\left( \dot{H}+3H^{2}\right) }+%
\frac{1}{2}\right) \right] dt\right\}.
\end{equation}

The condition of the smallness of the non-equilibrium bulk viscous pressure
with respect to the local equilibrium pressure can be checked from the
relation
\begin{equation}
\left| \frac{\Pi }{p}\right| =\left| \frac{\frac{\left( \gamma -2\right)
K^{2}}{2V^{2}}-3\gamma H^{2}-2\dot{H}}{3\left( \gamma -1\right) H^{2}\left(
1-\frac{K^{2}}{6V^{2}H^{2}}\right) }\right| = \left| \frac{\frac{\gamma -2}{2}A(t)-\gamma -\frac{2}{3H^{2}}\frac{dH}{dt}}{%
\left( \gamma -1\right) \left( 1-\frac{A(t)}{2}\right) }\right| 
\end{equation}

\end{document}